\documentclass[sn-mathphys,Numbered]{sn-jnl}

\usepackage{graphicx}%
\usepackage{multirow}%
\usepackage{amsmath,amssymb,amsfonts}%
\usepackage{amsthm}%
\usepackage{mathrsfs}%
\usepackage[title]{appendix}%
\usepackage{xcolor}%
\usepackage{textcomp}%
\usepackage{manyfoot}%
\usepackage{booktabs}%
\usepackage{algorithm}%
\usepackage{algorithmicx}%
\usepackage{algpseudocode}%
\usepackage{listings}%
\usepackage{subfigure}

\theoremstyle{thmstyleone}%
\newtheorem{theorem}{Theorem}
\newtheorem{lem}[theorem]{Lemma}%

\theoremstyle{thmstyletwo}%

\theoremstyle{thmstylethree}%
\newcommand{\sgn}{\operatorname{sgn}}

\newcommand{\rme}{\mathrm{e}}
\newcommand{\rmi}{\mathrm{i}}

\def\eqref#1{\textup{(\ref{#1})}}  
\newcommand{\eref}[1]{Eq.~\textup{(\ref{#1})}}

\newcommand{\sref}[1]{Sec.~\ref{#1}}

\newcommand{\cref}[1]{Conjecture~\ref{#1}}
\newcommand{\Cref}[1]{Conjecture~\ref{#1}}

\newcommand{\rcite}[1]{Ref.~\cite{#1}}

\begin{document}

\title[Article Title]{Efficient molecular conformation generation with quantum-inspired algorithm}

\author[1,2]{\fnm{Yunting} \sur{Li}}\email{liyt20@fudan.edu.cn}
\author[2]{\fnm{Xiaopeng} \sur{Cui}}\email{cuixiaopeng5@huawei.com}
\author[3]{\fnm{Zhaoping} \sur{Xiong}}\email{xiongzhp@shanghaitech.edu.cn}
\author[2]{\fnm{Zuoheng} \sur{Zou}}\email{zouzuoheng2@huawei.com}
\author[2]{\fnm{Bowen} \sur{Liu}}\email{liubowen35@huawei.com}
\author[2]{\fnm{Bi-Ying} \sur{Wang}}\email{biying@mail.ustc.edu.cn}
\author[2]{\fnm{Runqiu} \sur{Shu}}\email{1020288106@qq.com}
\author[1]{\fnm{Huangjun} \sur{Zhu}}\email{zhuhuangjun@fudan.edu.cn}
\author*[3]{\fnm{Nan} \sur{Qiao}}\email{qiaonan3@huawei.com}
\author*[2,4,5,6,7,8]{\fnm{Man-Hong} \sur{Yung}}\email{yung.manhong@huawei.com}

\affil[1]{\orgdiv{Institute for Nanoelectronic Devices and Quantum Computing}, \orgname{Fudan University}, \orgaddress{\city{Shanghai}, \postcode{200433}, \country{China}}}

\affil[2]{\orgdiv{Central Research Institute}, \orgname{Huawei Technologies}, \orgaddress{\city{Shenzhen}, \postcode{518129}, \country{China}}}

\affil[3]{\orgdiv{Laboratory of Health Intelligence}, \orgname{Huawei Cloud Computing Technologies Co., Ltd}, \orgaddress{\city{Guizhou}, \postcode{550025}, \country{China}}}

\affil[4]{\orgdiv{Shenzhen Institute for Quantum Science and Engineering}, \orgname{Huawei Cloud Computing Technologies Co., Ltd}, \orgaddress{\city{Guizhou}, \postcode{550025}, \country{China}}}

\affil[5]{\orgdiv{Laboratory of Health Intelligence}, \orgname{Southern University of Science and Technology}, \orgaddress{\city{Shenzhen}, \postcode{518055}, \country{China}}}

\affil[6]{\orgdiv{International Quantum Academy}, \orgaddress{ \city{Shenzhen}, \postcode{518048}, \country{China}}}

\affil[7]{\orgdiv{Guangdong Provincial Key Laboratory of Quantum Science and Engineering}, \orgname{Southern University of Science and Technology}, \orgaddress{\city{Shenzhen}, \postcode{518055}, \country{China}}}

\affil[8]{\orgdiv{Shenzhen Key Laboratory of Quantum Science and Engineering}, \orgname{Southern University of Science and Technology}, \orgaddress{\city{Shenzhen}, \postcode{518055}, \country{China}}}

\abstract{

\textbf{Context} 
Conformation generation, also known as molecular unfolding (MU), is a crucial step in structure-based drug design, remaining a challenging combinatorial optimization problem. 
Quantum annealing (QA) has shown great potential for solving certain combinatorial optimization problems over traditional classical methods such as simulated annealing (SA).  However, a recent study showed that a 2000-qubit QA hardware was still unable to outperform SA for the MU problem. Here, we propose the use of quantum-inspired algorithm to solve the MU problem, in order to go beyond traditional SA. We introduce a highly-compact phase encoding method which can exponentially reduce the representation space, compared with the previous one-hot encoding method. For benchmarking, we tested this new approach on the public QM9 dataset generated by density functional theory (DFT). The root-mean-square deviation between the conformation determined by our approach and DFT is negligible (less than about 0.5Å), which underpins the validity of our approach. Furthermore, the median time-to-target metric can be reduced by a factor of five compared to SA. Additionally, we demonstrate a simulation experiment by MindQuantum using quantum approximate optimization algorithm (QAOA) to reach optimal results. These results indicate that quantum-inspired algorithms can be applied to solve practical problems even before quantum hardware become mature.

\textbf{Methods} The objective function of MU is defined as the sum of all internal distances between atoms in the molecule, which is a high-order unconstrained binary optimization (HUBO) problem. The degree of freedom of variables are discretized and encoded with binary variables by the phase encoding method. We employ the quantum-inspired simulated bifurcation algorithm for optimization. The public QM9 dataset is generated by DFT. The simulation experiment of quantum computation is implemented by MindQuantum using QAOA.

}


\maketitle

\section{Introduction}

Efficient drug discovery is of significant interest in pharmaceutical science. Structure-based drug design with computer-assisted methods \cite{Leelananda2016,dosSantos2018,Vikram2020} has attempted to accelerate the discovery of drug candidates. 
The ability to accurately generate bioactive and low-energy conformations of small molecules from conformation generation is a theme of central importance in computational drug design \cite{Saunders1990,Wiley1991,Izrailev2006,Agrafiotis2007,Watts2010}. The conformation of a molecule greatly affects its binding behaviour and efficacy \cite{Copeland2011}, as drug receptors are highly sensitive to the structure of molecules binding to them \cite{STRUTHERS1985}. Many widely used techniques in drug discovery, such as molecular docking \cite{Daniel1996MD,Kitchen2004MD,Huang2006MD,Meng2011MD,Pagadala2017MD}, pharmacophore modeling \cite{Kristam2005,SCHWAB2010}, 3D-QSAR \cite{Verma2010} and virtual screening \cite{LYNE20021047,Stahura2005}, depend heavily on the ability to generate conformations close to the structure of the ligand in the protein-ligand complex of interest. Each inappropriate conformation increases the chance that the downstream applications will produce inaccurate results.
Existing conformation generation algorithms typically focus on minimizing the internal energy of the molecule with systematic methods and stochastic methods for optimization \cite{Izgorodina2007,Hawkins2010,OBoyle2011,Friedrich2017}. However, the computational resources needed are quite prohibitive due to the complex expression of the energy function and grow exponentially with the degrees of freedom.

In recent years, quantum computing has emerged as a promising tool to accelerate the computation and reduce resources by exploiting quantum mechanical systems. Thanks to the overall development of hardware and software of quantum computing, it has become widely used in many fields, including drug discovery \cite{Y2018IBM,Carlos2021,Irawan2022}. More practically, quantum-inspired algorithms \cite{Zhang2011,Rss2020,Arrazola2020quantuminspired,liu2020,Irawan2022} are currently accessible and effective in many applications, particularly for linear algebra problems \cite{Patrick2014SVM,Gilyen2018,Ewin2019,Arrazola2020quantuminspired,Ewin2021PCA,Gilyen2022improvedquantum} and combinatorial optimization problems \cite{Han2002,Edward2014QAOA,Tiunov2019,Hayato2016,Hayato2019,Hayato2019JPS,Hayato2021}.

Recently, Mato et al. have applied quantum annealing (QA) to the conformation generation problem \cite{Mato2022}.  
Since each atom prefers to repel each other and occupy enough space in the free vacuum environment, a simple and effective way to quantify the best configuration is to maximize the molecular volume, that is, the sum of all the internal distances between atoms in the molecule. This problem can be formulated as an combinatorial optimization problem with rotatable bonds within the molecules as variables. This distance-heuristic conformation generation is known as the \emph{molecular unfolding} (MU) problem.
Mato et al. developed a quantum MU approach and executed it on the latest QA hardware (D-wave advantage and 2000Q). However, QA did not outperform the simulated annealing (SA) algorithm \cite{SA1983,SA2015}.

Our research focuses on the MU problem, which can be categorized as an optimization problem. To address this issue, we explore the effectiveness of quantum-inspired algorithms for support. Specifically, we employ the \emph{simulated bifurcation} (SB) algorithm, which is a purely quantum-inspired algorithm based on quantum adiabatic optimization using nonlinear oscillators \cite{Hayato2016,Hayato2019,Hayato2019JPS,Hayato2021,Kanao2022,Dmitry2021}. Unlike the gradient method, SB is driven by Hamiltonian dynamics evolution, and allows for the simultaneous updating of variables, resulting in great parallelizability and accelerated combinatorial optimization. SB also shows significant potential for finding the global minima. Recently, two SB variants, namely ballistic SB (bSB) and discrete SB (dSB), have been proposed for further enhancement \cite{Hayato2021}.

In this work, we map the MU problem into a high-order unconstrained binary optimization (HUBO) problem. We propose an efficient phase encoding method to encode degrees of freedom of binary variables. The quantum-inspired algorithm bSB is applied to solve the HUBO problem with advantage of fast convergence and high parallelizability. Compared with the previous model in \rcite{Mato2022}, our approach reduces the number of binary variables exponentially. Moreover, we consider the well-known QM9 dataset \cite{Rama2014QM9,Ruddigkeit2012QM9} for testing. We show the performance of the volume ratio over iteration step window, and time-to-target metric. Compared with the famous SA algorithm, our approach has a significant advantage in finding a solution with higher volume ratio and faster convergence. Our approach can reduce the median time-to-target metric by a factor of five, which demonstrates its high efficiency. In addition, we show that the root-mean-square deviation (RMSD) between the unfolding conformation and the conformation determined by density functional theory (DFT) is less than about 0.5Å, which demonstrates the validity of our approach. Furthermore, we demonstrate a simulation experiment of MU by MindQuantum using the quantum approximate optimization algorithm (QAOA) to reach the optimal result

\section{Model and methods}\label{sec:methods}

\subsection{Molecular unfolding problem}\label{sec:MU problem}

Considering the repulsive interactions between electrons inside atoms, a reasonable configuration of a molecular is required to reach the maximal molecular volume. 
Given the 3D coordinates of atoms and the chemical bond properties of connection, the geometric shape of the ligand is obtained. 
The geometric description of a ligand corresponds to a graph $G=(V, E)$ in three dimensions. Each atom in the ligand corresponds to a vertex in the graph, while each chemical bond connecting two atoms corresponds to an edge with a fixed length.

\begin{figure}[tbp]
\centering
\subfigure[Schematic diagram of molecular fragments and unfolding.]{
\begin{minipage}{\linewidth}
\centering
\includegraphics[scale=0.45]{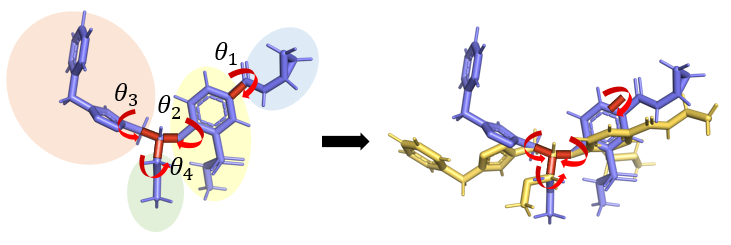}
\label{fig:demo}
\end{minipage}}
\subfigure[Schematic diagram of calculating the square distance between two atoms.]{
\begin{minipage}{\linewidth}
\centering
\includegraphics[scale=0.23]{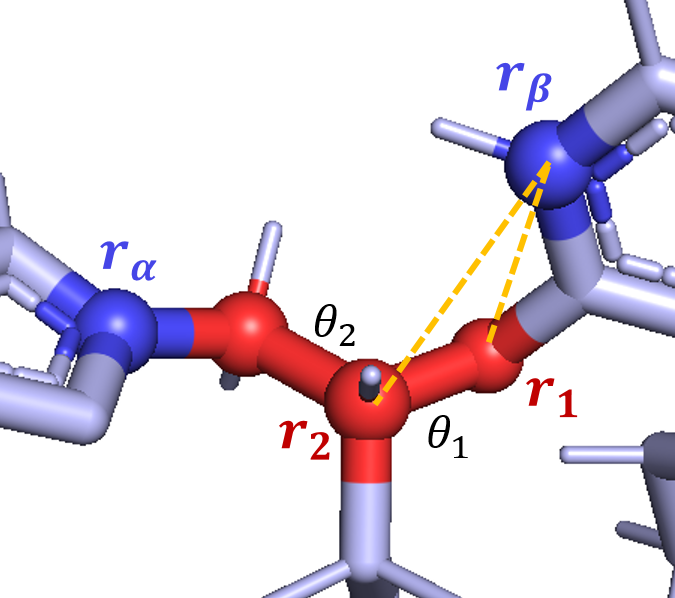}
\label{fig:RB}
\end{minipage}}
\caption{Schematic diagram of MU. (a) The red bonds are rotatable bonds. The color shaded area each includes a fragment. The rotatable bonds split the molecule into several disjoint fragments. The purple ligand is the given rough conformation. The yellow ligand is the unfolding conformation.
(b) The two atoms $\alpha, \beta$ are labeled as blue spheres with positions $\vec{r}_{\alpha}, \vec{r}_{\beta}$ respectively. The related rotabable bonds are colored red and the corresponding torsional angles $\theta_i$, $i=1,2$ are shown beside. One fixed atom on each rotatable bond $i$ is labeled as a red sphere with position $\vec{r}_i$. The orange dotted lines represent $\vec{r}_{\beta} - \vec{r}_i$ for each rotatable bond.}
\label{fig:MU}
\end{figure}

We show the schematic diagram of MU in Figure~\ref{fig:MU} .
While the bond lengths and bond angles are fixed, the freedom in the geometric description depends on a set of \emph{rotatable bonds}. Each rotatable bond splits the molecule into two nonempty disjoint \emph{fragments}. The whole molecular is split into several disjoint fragments. The number of rotatable bonds depends on the specific structure of the ligand under consideration.  
New configurations can be generated by rotating a fragment of a \emph{torsional angle} around a connected rotatable bond. 
The torsional angles of each rotatable bond determines the configuration of the ligand. Note that the rotation along the rotatable bonds will not effect the bond length and bond angles.
As shown in Figure~\ref{fig:demo}, the four red bonds are the chosen rotatable bonds. The four color shaded areas are four fragments separated by the rotatable bonds. The purple ligand is the given conformation and the yellow ligand is the unfolding conformation.

Suppose the ligand has $M$ rotatable bonds. Denote by $\theta_i \in [0, 2\pi), \ i=1,2,\dots,M$ the corresponding $M$ torsional angles and $\Theta={(\theta_1, \dots, \theta_M)}$ the compact notation of all angles. The rotation operator along a rotatable bond reads
\begin{align}
R(\theta)=I +\sin \theta K + (1-\cos \theta) K^2,
\end{align}
where $K$ depends on the unit vector along the rotation axis $\vec{k}=(k_x, k_y,k_z)$
\begin{align}
K=
\begin{pmatrix}
0 & -k_z & k_y \\
k_z & 0 & -k_x \\
-k_y & k_x & 0
\end{pmatrix}.
\end{align}
Consider two atoms $\alpha, \beta$ in the ligand with initial positions $\vec{r}_{\alpha}, \vec{r}_{\beta}$. Then the initial square distance between the two atoms reads
\begin{align}
D_{\alpha \beta}^{\mathrm{ini}} = \lVert \vec{r}_{\alpha} - \vec{r}_{\beta} \rVert ^2.
\end{align}

To clarify the dependence of the square distance on the torsional angles, suppose two atoms are in different fragments and thus affected by rotatable bonds.
For each pair of these two atoms, the related rotatable bonds are those appearing in the shortest path connecting them. 
Given two atoms $\alpha, \beta$, the schematic diagram of calculating the square distance is shown in Figure~\ref{fig:RB}. The related rotatable bonds are colored red. The corresponding torsional angles along the shortest path are denoted as $\Theta^{\alpha, \beta} = (\theta_1, \dots, \theta_m), m\le M $. Here we have $m=2$ in Figure~\ref{fig:RB}.
Without loss of generality, we assume that only the atom $\beta$ is rotated along the rotatable bonds in sequence. Then the updated square distance according to the corresponding torsional angles reads
\begin{align}
D_{\alpha \beta} (\Theta^{\alpha, \beta}) = \lVert \vec{r}_{\alpha} - \mathbf{R}(\vec{r}_{\beta}; \Theta^{\alpha, \beta}) \rVert ^2,
\end{align}
where $\mathbf{R}(\vec{r}_{\beta}; \Theta^{\alpha, \beta})$ maps $\vec{r}_{\beta}$ to the rotated position.
To be specific, for atoms $\alpha, \beta$ we have
\begin{align}\label{eq:rotation all}
\mathbf{R}(\vec{r}_{\beta}; \Theta^{\alpha, \beta}) =\mathbf{R}_m \circ \cdots \circ \mathbf{R}_2 \circ \mathbf{R}_1 (\vec{r}_{\beta}),
\end{align}
and for each rotation operation $\mathbf{R}_i, i=1, \dots, m$ we have 
\begin{align}\label{eq:rotation one}
\mathbf{R}_i(\vec{r}_{\beta}) = R(\theta_i ) (\vec{r}_{\beta} - \vec{r}_i) + \vec{r}_i,
\end{align}
where $\theta_i$ is the corresponding torsional angle of rotatable bond $i$, and $\vec{r}_i$ is the position of one fixed atom on the corresponding rotatable bond, labeled as a red sphere in Figure~\ref{fig:RB}.

The molecular volume is defined as the total sum of square distances between each pair of atoms in the ligand as follows \cite{Mato2022}:
\begin{align}
D(\Theta) = \sum_{\alpha,\beta \in V, \alpha \ne \beta} D_{\alpha \beta} (\Theta^{\alpha, \beta}).
\end{align}
Alternatively, we can focus on those terms in which $\alpha$ and $\beta$ belong to different fragments because other terms do not depend on the torsional angles and are thus irrelevant to the MU problem. Then the molecular volume can be expressed as follows:
\begin{align}\label{eq:molecular volume}
D(\Theta) = \sum_{\alpha \in V_i, \beta \in V_j, i \ne j} D_{\alpha \beta} (\Theta^{\alpha, \beta}),
\end{align}
where $V_i, V_j$ are subsets of vertexes which correspond to fragments of the ligand. Note that $V_i \cap V_j = \varnothing$ and $\sum_i V_i=V$.

Now the MU problem reduces to an optimization problem: find an ensemble of torsional angles to maximize the molecular volume $D(\Theta)$ defined in \eref{eq:molecular volume}.

\subsection{HUBO formulation in phase encoding}\label{sec:HUBO form}

The objective function can be transformed into a \emph{High-order Unconstrained Binary Optimization (HUBO)} formulation. We discretize the torsional angles into some uniform discrete values and introduce binary variables for encoding. 
Here we propose an efficient \emph{phase encoding}, which exponentially reduces the number of binary variables required for encoding each torsional angle compared with the traditional one-hot encoding \cite{Mato2022}.

Suppose the value of each torsional angle $\theta_i$ is chosen from $d$ discrete values $\{\phi_0, \phi_1, \dots, \phi_{d-1} \}$ which are uniformly distributed from 0 to $2\pi$,
\begin{align}\label{eq:discrete values}
\phi_k =k \frac{2 \pi}{d},\quad k=0,1,\dots, d-1.
\end{align}
Then the trigonometric functions can be discretized accordingly $\{ \sin(\phi_0), \sin(\phi_1), \dots, \sin(\phi_{d-1})\}$ (same for $\cos$).

In the one-hot encoding method, each torsional angle $\theta_i$ corresponds to $d$ binary variables $\{b_{ik}\}_{k=0}^{d-1}$ with ${ b_{ik} \in \{0,1\}}$. $b_{ik}=1$ if $\theta_i=\phi_k$; otherwise $b_{ik}=0$. The detail of one-hot encoding is elaborated in Sec.~S2 in the Supplemental Material \cite{supp}.
Note that the number of binary variables needed is $Md$, where $M$ is the number of rotatable bonds, and the number of HUBO terms for each $\sin(\phi_k)$ and $\cos(\phi_k)$ is $d$, which leads to a large number of variables and terms in the objective function.
The resource for one-hot encoding is quite expensive.

In the phase encoding method, the $d$ discrete values $\{\phi_k\}_{k=0}^{d-1}$ in \eref{eq:discrete values} can be encoded with only $n=\log_2 d$ binary variables, assuming that $d$ is a power of 2.
Then each torsional angle $\theta_i$ corresponds to $n$ binary variables $\{b_{ij}\}_{j=0}^{n-1}$ with ${b_{ij} \in \{\pm 1\}}$. 
A key technique to realize the phase encoding is shown in the following lemma from \rcite{PE2023}.

\begin{lem}
Suppose $n$ is a positive integer and $\{\psi_k\}_{k=0}^{2^n-1}$ are phases uniformly distributed on the circle. 
Then we can construct a polynomial $p_n(\mathbf{s})$ with $2^{n-1}$ terms such that $\{p_n(\mathbf{s})|\mathbf{s}\in \{\pm 1\}^n\}=\{\rme^{\rmi \psi_k}\}_{k=0}^{2^n-1}$.
\end{lem}

For $n=2$ and $\psi_k = \frac{2\pi}{2^n} k = \frac{\pi}{2} k$, we have
\begin{equation}
p_2(\mathbf{s}) = c_0 s_0 + c_1 s_1, 
\end{equation}
where $\mathbf{s}=(s_0, s_1) \in \{\pm 1\}^2$ and the coefficients are
\begin{equation}
\begin{pmatrix}
c_0 \\ c_1
\end{pmatrix}
= \frac{1}{2}
\begin{pmatrix}
1-\rmi \\ 1+\rmi
\end{pmatrix}.
\end{equation}
The value of $k$ has a unique correspondence with the values of binary variables. For example, $\psi_1 = \frac{\pi}{2}$ corresponds to $s_0=-1, s_1=1$. Other possibilities can be similarly calculated.

\begin{figure}[t]
\centering
\includegraphics[scale=0.28]{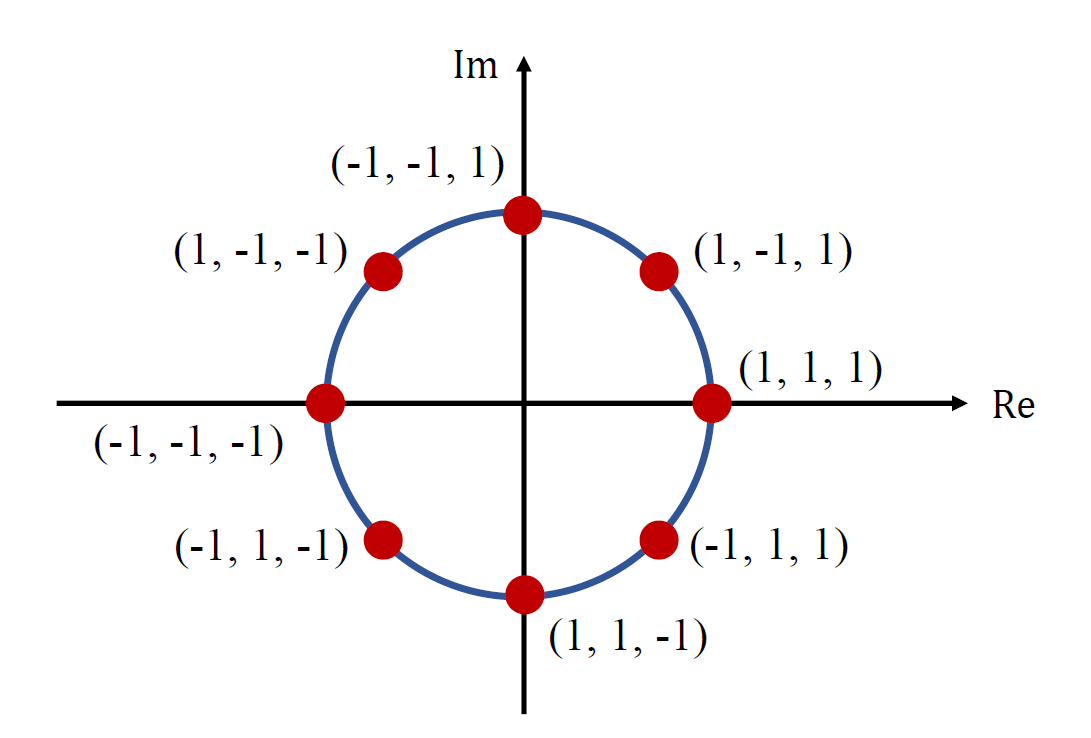}
\caption{The correspondence between binary variables and discrete phases on the circle in phase encoding with $n=3$. The red points represent uniformly distributed phases on the circle. The value of $(s_0, s_1, s_2)$ for each phase is shown beside the point.}
\label{fig:phase_encoding}
\end{figure}

For $n=3$ and $\psi_k =\frac{2\pi}{2^n} k = \frac{\pi}{4} k$, we have
\begin{equation}
p_3(\mathbf{s}) = c_0 s_0 + c_1 s_1 + c_2 s_2 + c_3 s_0 s_1 s_2, 
\end{equation}
where $\mathbf{s}=(s_0, s_1, s_2) \in \{\pm 1\}^3$ and the coefficients are
\begin{equation}
\begin{pmatrix}
c_0 \\ c_1 \\ c_2 \\ c_3
\end{pmatrix}
=\frac{1}{4}
\begin{pmatrix}
1+(\sqrt{2}-1)\rmi \\ 1-(\sqrt{2}+1)\rmi \\ (1+\sqrt{2})+\rmi \\ (1-\sqrt{2})+\rmi
\end{pmatrix}.
\end{equation}
The correspondence between binary variables and discrete phases on the circle is shown in Figure~\ref{fig:phase_encoding}. The red points represent phases uniformly distributed on the circle. The value of $(s_0, s_1, s_2)$ for each phase is shown beside the point.
Similarly, for any value of $n$, we can obtain the polynomial expression of discrete phases from \rcite{PE2023}.

Note that the torsional angle appears in the rotation operator as $\sin(\theta_i), \cos(\theta_i)$ and $\theta_i \in \{\phi_k\}_k$. We propose the efficient polynomial expressions of $\sin(\theta_i), \cos(\theta_i)$ from the polynomial expression $p_n(\mathbf{b}_i)$ of $\rme^{\rmi \phi_k}$ read
\begin{align}
\sin(\theta_i) = \Im p_n(\mathbf{b}_i), \quad \cos(\theta_i) = \Re p_n(\mathbf{b}_i),
\end{align}
where $\mathbf{b}_i=(b_{i0},\dots, b_{in-1}) \in \{\pm 1\}^n$. 

Therefore, by virtue of the phase encoding, the objective function can be directly written as a HUBO formulation as follows:
\begin{align}
O(\{b_{ij} \}) = - D(\Theta).
\end{align}
Then the MU problem reduces to minimizing the objective function $O(\{b_{ij} \})$.

Note that the number of binary variables needed is only $M \log_2 d$, which is exponentially reduced from one-hot encoding with respect to $d$. The number of HUBO terms for each $\sin(\phi_k)$ or $\cos(\phi_k)$ is only $2^{n-1}=d/2$, compared to $d$ in the one-hot encoding, which leads to much fewer terms in the objective function.

Here we show an example of an objective function in HUBO formulation of a molecule in the QM9 dataset we considered in \sref{sec:dataset} with $M=2$ rotatable bonds and $n=3$ binary variables for each bond in \eref{eq:HUBO eg}.

\begin{align}\label{eq:HUBO eg}
O(\{b_{ij} \}) =& -240.7234-0.0112b_{00}b_{12}+0.0378b_{10}b_{11}b_{12} +0.0049b_{10}b_{00}b_{11}b_{01}b_{12}b_{02} \\ \nonumber
&+0.0003b_{00}b_{01}b_{02}
+0.0004b_{00}b_{11}+0.0286b_{11}b_{01}-0.0115b_{10}b_{02} \\ \nonumber
&-0.0118b_{00}b_{11}b_{01}b_{02}+0.0001b_{00}b_{01}b_{12}b_{02}-0.027b_{12}b_{02}\\ \nonumber
&-0.0008b_{10}b_{11}b_{12}b_{02}-0.0003b_{10}b_{00}b_{11}b_{12}
-0.0119b_{10}b_{11}b_{01}b_{12}\\ \nonumber
&-0.0003b_{01}b_{12}-0.0833b_{11}-0.0048b_{10}b_{00}+0.001b_{11}b_{02}-0.0002b_{10}b_{01} \\ \nonumber
&-0.0008b_{01}+0.2475b_{10}+0.0005b_{10}b_{00}b_{12}
-0.0025b_{00}+0.0001b_{10}b_{01}b_{12}\\ \nonumber
&+0.0012b_{10}b_{12}b_{02}-0.0112b_{10}b_{12}+0.5896b_{12}-0.0059b_{02}.
\end{align}

\subsection{Quantum-inspired algorithm: simulated bifurcation}\label{sec:SB}

Simulated bifurcation (SB) \cite{Hayato2016,Hayato2019,Hayato2019JPS,Hayato2021} is a heuristic technique for accelerating combinatorial optimization. It was first proposed for solving the Ising problem of finding a spin configuration minimizing the Ising energy
\begin{align}
E_{\mathrm{Ising}} = - \frac{1}{2} \sum_{i=1}^N \sum_{j=1}^N J_{ij}s_i s_j,
\end{align}
where $N$ is the number of spins, $s_i$ denotes the $i$th spin of value $+1$ or $-1$, and $J_{ij}$ is the coupling coefficient between the $i$th and $j$th spins. Note that the Ising energy is formulated as a \emph{Quadratic-order Unconstrained Binary Optimization} (QUBO) problem.

To solve the Ising problem, a network of quantum-mechanical nonlinear oscillators \emph{Kerr-nonlinear parametric oscillators} (KPOs) is used to construct the quantum Hamiltonian \cite{Hayato2016,Hayato2019}.
After deriving the classical Hamiltonian from the quantum Hamiltonian, we can obtain the equations of motion of position $x_i$ and momentum $y_i$ of a particle corresponding to the $i$th spin from Hamilton-Jacobi equation.

Later, to avoid errors from using continuous variables $x_i$ instead of discrete variables $s_i$, a variant \emph{ballistic SB} (bSB) was introduced \cite{Hayato2021}. Inelastic walls are set at $x_i=\pm 1$, that is, if $|x_i|>1$ during the evolution, replace $x_i$ with $\sgn(x_i)$ and set $y_i=0$. The equations of motion of bSB read \cite{Hayato2021}
\begin{equation}\label{eq:EOM bSB}
\begin{aligned}
\dot{x}_i &= \frac{\partial H_{\mathrm{bSB}}}{\partial y_i} = a_0 y_i,  \\
\dot{y}_i &= -\frac{\partial H_{\mathrm{bSB}}}{\partial x_i} = \left[a_0 -a(t)\right] x_i + c_0 \sum_j^N J_{ij} x_j,
\end{aligned}
\end{equation}
where $x_i, y_i$ are the position and momentum of a particle corresponding to the $i$th spin, $a(t)$ is a control parameter increasing from 0 to $a_0$, and $a_0, c_0$ are positive constants.
The Hamiltonian $H_{\mathrm{bSB}}$ reads 
\begin{equation}
H_{\mathrm{bSB}} = \frac{a_0}{2} \sum_i^N y_i^2 + V_{\mathrm{bSB}},
\end{equation}
and if all $|x_i| \le 1 $ the potential $V_{\mathrm{bSB}}$ reads
\begin{equation}
V_{\mathrm{bSB}} =
\frac{a_0-a(t)}{2} \sum_i^N x_i^2  - \frac{c_0}{2} \sum_i^N \sum_j^N J_{ij}x_i x_j,
\end{equation}
otherwise it reads infinity.
All positions and momentums are randomly set around zero at the initial time. When $a(t)$ becomes sufficiently large, multiple local minimas of the potential will appear in form of bifurcations. Then the particles will adiabatically follow one of the minima driven by the Ising energy in the Hamiltonian. Finally, $\sgn(x_i)$ gives the solution of the $i$th spin.

Fortunately, recent study has shown that the equations of motion of bSB can be generalized to HUBO formulation \cite{Kanao2022,Dmitry2021}. By replacing the Ising energy term in $V_{\mathrm{bSB}}$ with a general higher-order objective function $E(\vec{x})$, we have the general equaitons of motion as follows:
\begin{equation}
\begin{gathered}
\dot{x}_i = \frac{\partial H_{\mathrm{bSB}}}{\partial y_i} = a_0 y_i, \\
\dot{y}_i = -\frac{\partial H_{\mathrm{bSB}}}{\partial x_i} = \left[a_0 -a(t)\right] x_i + c_0 \frac{\partial E(\vec{x})}{\partial x_i},
\end{gathered}
\end{equation}
where $\vec{x}$ is the vector of positions of all spins. This generalization brings bSB algorithm to a brand new stage, which not only performs better than the second-order one but also avoids the QUBO requirement and yields more application scenarios.

\section{Results and discussion}\label{sec:result}

To demonstrate the performance of our approach, here we conduct a comparative analysis between bSB with phase encoding and SA with one-hot encoding, examining various aspects, such as the volume ratio over iteration steps and the time-to-target metric. The QM9 dataset \cite{Rama2014QM9,Ruddigkeit2012QM9} is employed for concrete comparison. 

\subsection{Dataset}\label{sec:dataset}

The QM9 (Quantum Machine 9) dataset consists of 133,885 species with up to nine heavy atoms (C, O, N and F) out of the GDB-17 chemical universe of 166 billion organic molecules \cite{Rama2014QM9,Ruddigkeit2012QM9}. This dataset provides quantum chemical properties for a relevant, consistent, and comprehensive chemical space of small organic molecules. The conformations in this dataset are generated by DFT (SOTA). Hence this dataset may serve the benchmarking and development of research with great influence and generality.

\begin{figure}[htbp]
\centering
\includegraphics[scale=0.8]{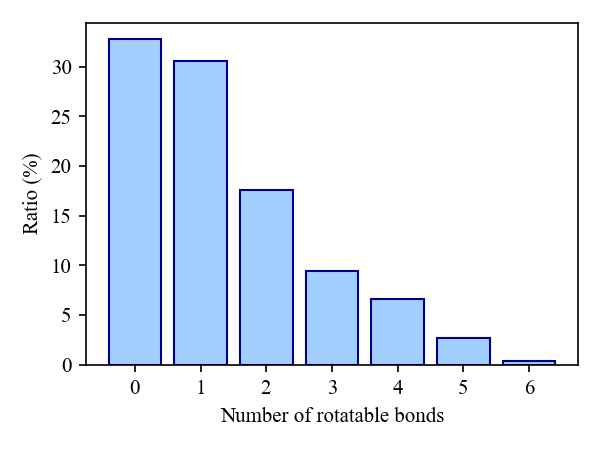}
\caption{The ratio of molecules with varying number of rotatable bonds in the QM9.}
\label{fig:RB ratio}
\end{figure}

The ratio of molecules in the QM9 with varying number of rotatable bonds is shown in Figure~\ref{fig:RB ratio}. The number of rotatable bonds is up to six. 
Therefore, within our study, we consider molecules with two to five rotatable bonds and randomly select 50 molecules for each number of rotatable bonds as testing dataset.

\subsection{Resources estimation}

Here we assume that each torsional angle is chosen from $d=16$ discrete values $\phi_k = \frac{\pi}{8} k, {k=0,1,\dots,15}$. The torsional interval of $\pi/8$ is small enough to reach an acceptable unfolding quality.

\begin{figure}[tbp]
\centering
\includegraphics[scale=0.8]{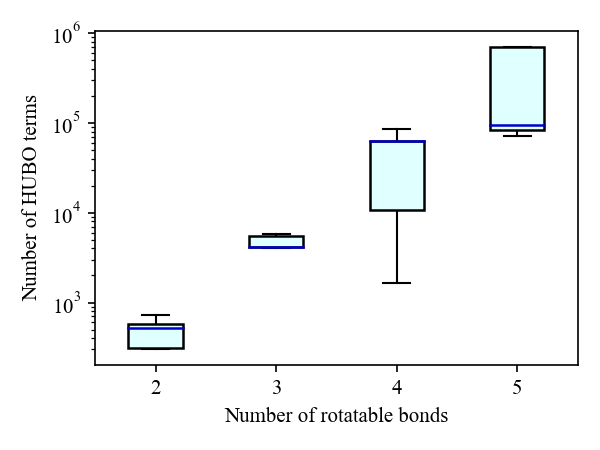}
\caption{The boxplot of the number of HUBO terms for varying number of rotatable bonds of the testing dataset.}
\label{fig:HUBO term}
\end{figure}

Suppose the number of rotatable bonds is $2 \le M\le 5$. The number of binary variables for each torsional angle is $n=4$, and the total number of binary variables is $8 \le nM \le 20$.
Further, we show the boxplot of the number of HUBO terms for varying number of rotatable bonds of the testing dataset in Figure~\ref{fig:HUBO term}. In a boxplot, the top and bottom lines (whiskers) indicate the maximum and minimum; the top and bottom of the box indicate the first quatile and third quatile; the middle line in the box indicates the median.
The median number of HUBO terms grows exponentially with the increase of the number of rotatable bonds. The distribution for each case depends on the structures of the randomly-selected molecule set.

\subsection{RMSD verification}

We show the root mean square deviation (RMSD) between the unfolding conformation and the conformation determined by density functional theory (DFT) in QM9 dataset and to verify the validity of the approach.

\begin{figure}[t]
\centering
\subfigure[The boxplot of RMSD determined by different methods.]{
\begin{minipage}{\linewidth}
\centering
\includegraphics[scale=0.75]{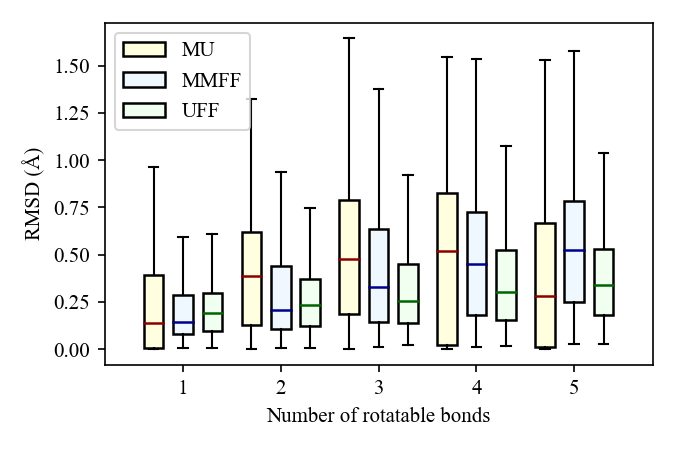}
\label{fig:RMSD}
\end{minipage}}
\subfigure[Better situation of MU.]{
\begin{minipage}{\linewidth}
\centering
\includegraphics[scale=0.4]{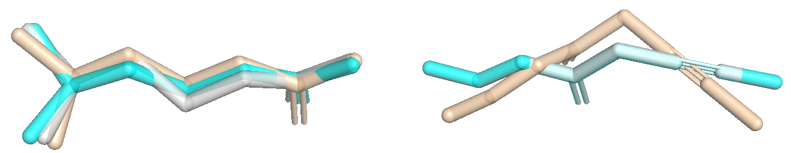}
\label{fig:QM9 demo better}
\end{minipage}}
\subfigure[Worse situation of MU.]{
\begin{minipage}{\linewidth}
\centering
\includegraphics[scale=0.4]{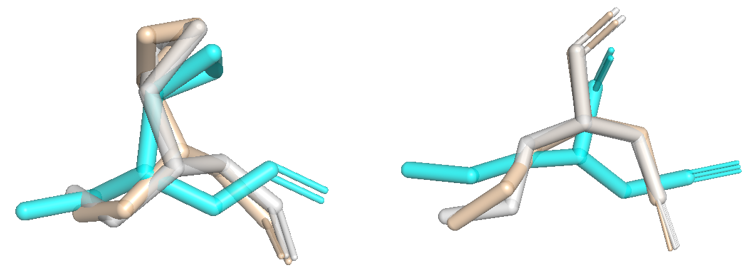}
\label{fig:QM9 demo worse}
\end{minipage}}
\caption{Comparison of RMSD between the conformations determined by DFT in QM9 dataset and from MU optimization, MMFF and UFF optimization in rdkit. 
(a) RMSD basically increases with the number of rotatable bonds, while for five rotatable bonds the RMSD of MU is smaller than MMFF and UFF.
(b) Better situation: The RMSD of MU is smaller than UFF. The molecular conformation of QM9 is colored in gray, while the optimized molecular conformation from MU is colored in cyan, and the one from UFF is colored in wheat.
(c) Worse situation: The RMSD of MU is larger than UFF.}
\label{fig:QM9 model}
\end{figure}

Here we conduct a random sampling of 2500 molecules each for two to four rotatable bonds, and 1000 molecules for five rotatable bonds. The unfolding conformation is from the brute-force search. 
The definition of RMSD is the average distance between atoms of two conformations as follows:
\begin{equation}
\mathrm{RMSD}=\sqrt{\frac{1}{N} \sum_{i=1}^N \lVert \vec{r}_i - \vec{r}_i' \rVert^2},
\end{equation}
where $N$ is the number of atoms in the molecule, $\vec{r}_i$ is the position vector of the $i$th atom in the first conformation and $\vec{r}_i'$ is the position vector of the same atom in the second conformation.
We show a boxplot of the RMSD between the conformations determined by DFT in QM9 dataset and from MU optimization, MMFF (Merck Molecular Force Field) and UFF (Universal Force Field) optimization in rdkit for varying numbers of rotatable bonds in Figure~\ref{fig:RMSD}.
As illustration, we show the molecular conformations optimized from MU and UFF, two demo molecules for better situation of MU in Figure~\ref{fig:QM9 demo better} and two for worse situation in Figure~\ref{fig:QM9 demo worse}. The molecular conformation of QM9 is colored in gray, while the optimized molecular conformation from MU is colored in cyan, and the one from UFF is colored in wheat. 
The comparison of the values of the molecular volume and RMSD is shown in Sec.~S3 in the Supplemental Material \cite{supp}.

We found that the RMSD generally increases with the number of rotatable bonds, but for five rotatable bonds, the RMSD from MU is smaller than that from MMFF and UFF. This observation suggests that as the complexity of the molecule increases, it becomes more challenging for force-field optimization to yield a good conformation.
Fortunately, we discovered that it is both reasonable and more efficient to consider the configuration with maximal molecule volume as the most probable one. This greatly simplifies the conformation generation process.

\subsection{Compared algorithms}

We compare the performance between SA with one-hot encoding (SA+OHE) in \rcite{Mato2022} and bSB with phase encoding (bSB+PE).
The brute-force search is used to give the optimal solutions.

SA is a meta-heuristic algorithm useful in searching optimum in large solution spaces. The optimization process starts with a random initial configuration of the binary bits. Each successive step is to pick a neighbouring configuration and estimate the cost function. If the neighbour has a lower cost, then accept it as the current solution. Otherwise, accept it with an acceptance probability related to the simulated temperature. The binary variables associated with valid neighbours need to satisfy the constraint one-hot encoding. The initial temperature is set to yield an acceptance probability of 0.8 at the beginning. Then the temperature decreases geometrically following a cooling schedule.

\begin{figure*}[htbp]
\centering
\includegraphics[scale=0.65]{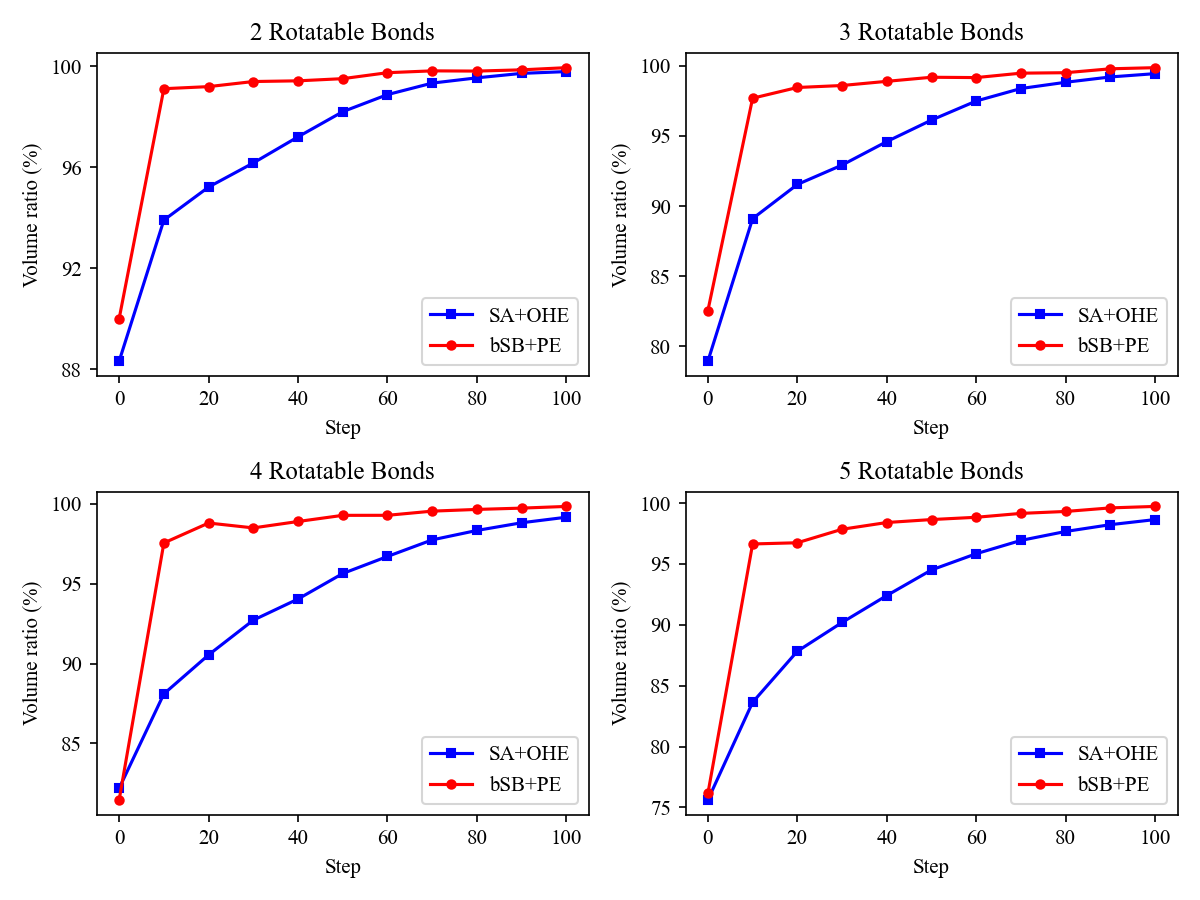}
\caption{Comparison of average volume ratio of QM9 dataset during 100 iteration steps for two to five rotatable bonds. Each point on the figure represents the average volume ratio over 20 samples for each molecule and 50 molecules at the current step.}
\label{fig:accu-step}
\end{figure*}

\subsection{Volume ratio comparison}

We compare the volume ratio over iteration steps obtained by different algorithms with an increasing number of rotatable bonds. 
The volume ratio $\alpha$ of each algorithm is defined as follows:
\begin{equation}
\alpha = \frac{O}{O_{\max}},
\end{equation}
where $O$ is the molecular volume reached by the optimization algorithm and $O_{\max}$ is the optimal solution from the brute-force search. 

We show the comparison on the average volume ratio during 100 iteration steps for two to five rotatable bonds in Figure~\ref{fig:accu-step}. Each point on the figure represents the average volume ratio over 20 samples for each molecule and 50 molecules at the current step. The solutions from each sample of a molecule may be different due to the random initialization. This comparison shows the convergence behaviour over iteration steps.
bSB has a faster convergence than SA at the beginning and has about 1\% higher volume ratio at the end of iteration window. 
Due to the definition of the volume ratio, the optimal solution $O_{\max}$ is a large value for most molecules, so a 1\% improvement still means a lot.

\begin{figure*}[htpb]
\centering
\includegraphics[scale=0.65]{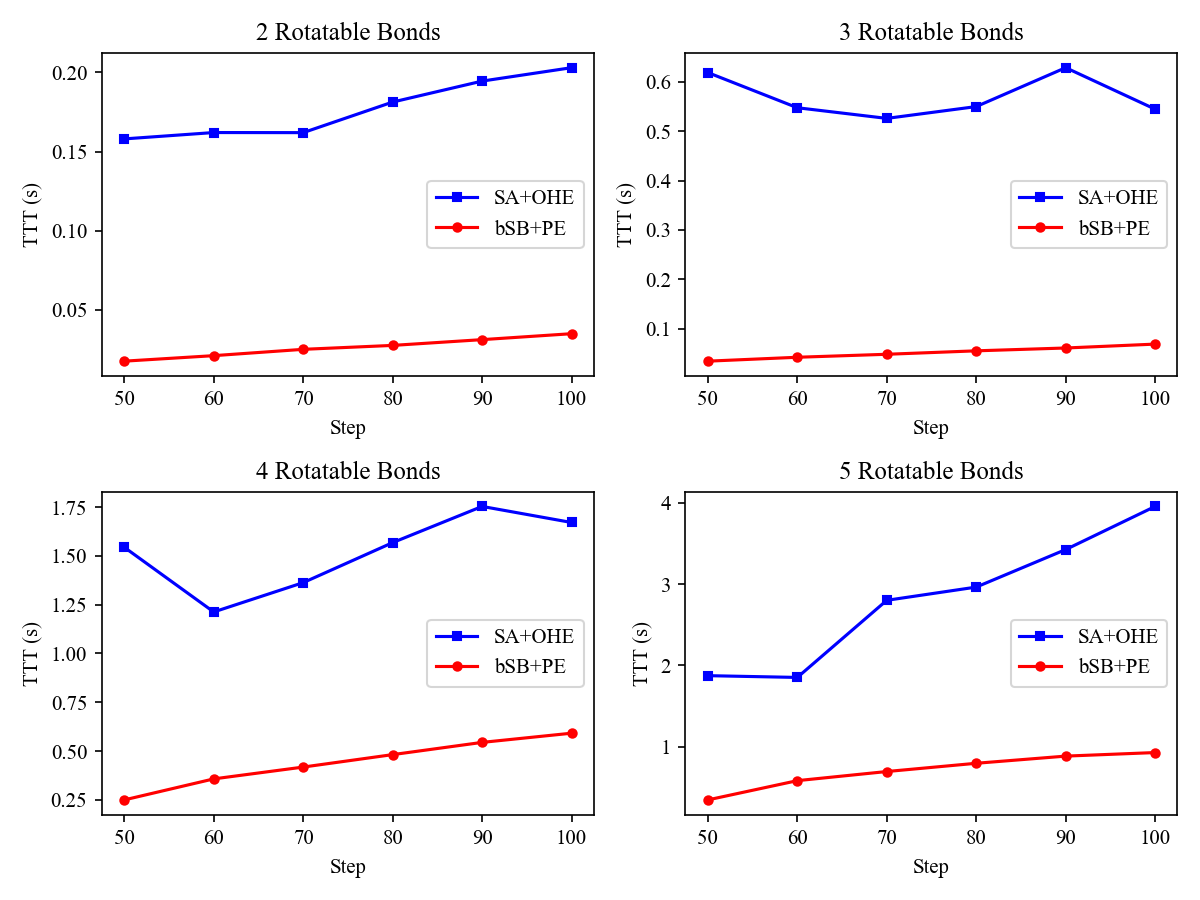}
\caption{Comparison of median TTT of QM9 dataset on varying iteration step windows. Each point on the figure represents the median TTT requied on the corresponding iteration step window over 50 molecules.}
\label{fig:TTT-step}
\end{figure*}

\subsection{Time-to-target comparison}

We compare different algorithms with respect to the time-to-target (TTT) metric for an increasing number of rotatable bonds.
The target here refers to a suboptimal solution which is defined as 99.7\% of the optimal volume. 
For SA and bSB, the TTT is defined as follows:
\begin{align}\label{eq:TTT}
\mathrm{TTT} = 
\begin{cases}
\frac{T}{N} \frac{\log(1-0.99)}{\log(1-p)}, \quad & \mathrm{if\ } 0 \le p <1, \\
\frac{T}{N}, \quad &\mathrm{if\ } p=1,
\end{cases}
\end{align}
where $T$ is the total execution time, $N$ is the sample number and $p$ is the success probability of reaching the target among $N$ samples. Better solvers have lower values of TTT.

We show the comparison of median TTT on varying iteration step windows in Figure~\ref{fig:TTT-step}. Each point on the figure represents the median TTT required on the corresponding iteration step window over $N=20$ samples for each molecule and 50 molecules. The solutions from each sample of a molecule may be different due to the random initialization, and the corresponding $p$ is obtained for each molecule on each iteration step window.
We find that the TTT of bSB is much shorter, and the TTT of SA is about 5 times of bSB at 100 iteration steps.
So bSB can achieve a higher volume ratio with a much shorter solution time.

In addition, the advantage of bSB in TTT is expected to be more significant when the number of rotatable bonds is larger. In Sec.~S4 in the Supplemental Material \cite{supp}, we use the same dataset as in \rcite{Mato2022} and show greater advantage for an increasing number of rotatable bonds.

\subsection{Quantum numerical simulation}\label{sec:simulation}

To show the applicability of our method on quantum computers, we demonstrate a simulation experiment using quantum approximate optimization algorithm (QAOA) \cite{farhi2014quantum}. All the simulation results are obtained with MindQuantum \cite{mq_2021}, which is a very powerful and efficient software library for quantum computation. 
MindQuantum provides basic quantum computation elements including quantum gate, quantum circuits, Hamiltonian, and quantum simulators; applications in noisy quantum simulation, quantum circuit compilation, qubit mapping, and other scenarios that are closer to real quantum chip environments; NISQ (Noisy Intermediate-Scale Quantum) algorithms including variational quantum algorithms and hybrid quantum-classical algorithms.

\begin{figure}[htpb]
\centering
\includegraphics[scale=0.6]{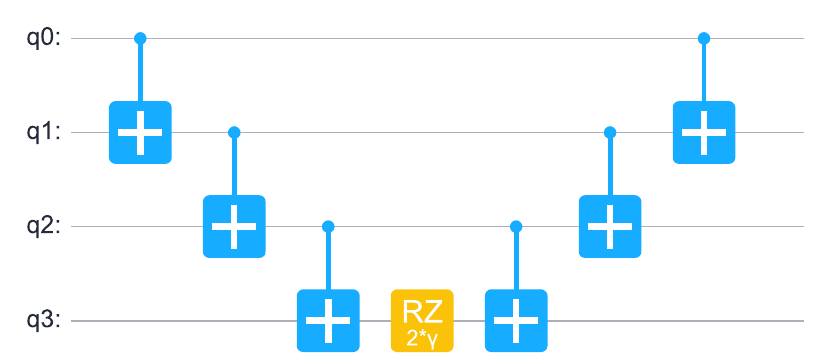}
\caption{The quantum circuit decomposition of $e^{\rmi \gamma Z_0Z_1Z_2Z_3}$ of the four-order term $Z_0Z_1Z_2Z_3$. $\gamma$ is a parameter in QAOA. $Z_0,Z_1,Z_2,Z_3=0,1$ are four binary variables.}
\label{fig:hubo_decompose}
\end{figure}

\begin{figure}[htbp]
\centering
\subfigure[The landscape of the expectation value of the original Hamiltonian.]{
\begin{minipage}{\linewidth}
\centering
\includegraphics[scale=0.42]{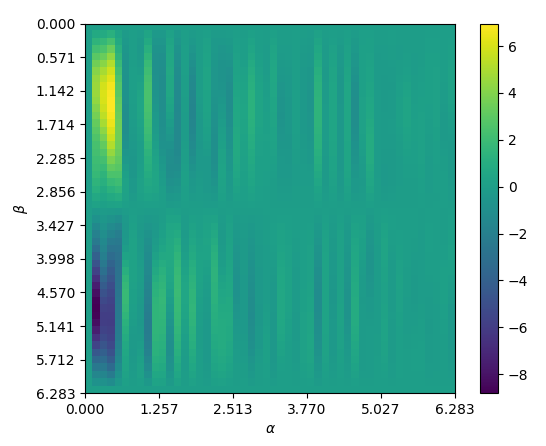}
\label{fig:landscape_initial}
\end{minipage}}
\subfigure[The landscape of the expectation value of the Hamiltonian after parameter rescaling.]{
\begin{minipage}{\linewidth}
\centering
\includegraphics[scale=0.42]{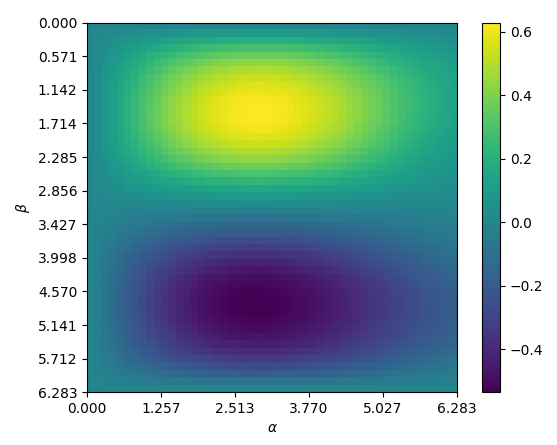}
\label{fig:landscape_rescale}
\end{minipage}}
\caption{The landscape of the expectation value of the Hamiltonian. $\alpha,\beta \in [0, 2\pi]$ are parameters of the phase layer and the mix layer in QAOA, respectively.}
\label{fig:landscape}
\end{figure}

QAOA is one of the most famous variational quantum algorithms (VQA) introduced to approximately solve combinatorial optimization problems. The ansatz of QAOA is inspired by the trotterization of the quantum adiabatic algorithm (QAA)~\cite{farhi2000quantum}. A single QAOA layer comprises one cost and one mixer layer, which can be further stacked to build a deeper circuit with more layers. For simplicity, we only consider the single layer QAOA with $p=1$ in this work.

Given an objective function in HUBO formulation, it is transformed to a quantum Hamiltonian and embedded into the cost layer, which is just the time evolution circuit of the whole Hamiltonian multiplying the parameter. 
Since the number of HUBO terms is usually too large to be realized on the quantum computers and many terms have extremely small coefficients, a simplification is introduced to eliminate the HUBO terms with coefficients less than a certain threshold value.

The cost layer of the QAOA circuit is composed of lots of multi-qubit rotation gates with each one corresponding to a high-order term in the Hamiltonian. 
For executing it on the quantum chip, the multi-qubit rotation gates should be decomposed into CNOT gates and rotational gates.
As shown in Figure~\ref{fig:hubo_decompose}, the evolution $e^{\rmi \gamma Z_0Z_1Z_2Z_3}$ of the four-order term $Z_0Z_1Z_2Z_3$ is decomposed into several basic quantum gates, where $\gamma$ is a parameter in QAOA and $Z_0,Z_1,Z_2,Z_3=0,1$ are four binary variables.
After preparing the parameterized quantum circuit, we can use classical optimizers to find optimized parameters which minimize the expectation of the HUBO Hamiltonian.

In this work, we randomly choose one molecule in the QM9 dataset for demonstration. No. 1282 molecule has three rotatable bonds each encoded by four binary variables, which results in 12 qubits. The number of HUBO terms is reduced from 811 to 76 by the coefficient threshold 0.1.
After the selection process, the remaining coefficient distribution of the HUBO terms is extremely broad, making optimization very challenging.
The landscape of the expectation value of the Hamiltonian is shown in Figure~\ref{fig:landscape_initial} with parameter $\alpha \in [0, 2\pi]$ of the phase layer and parameter $\beta\in [0, 2\pi]$ of the mixer layer in QAOA.
The minimum area is small and sharp, which is easy to fall into a local minima with randomly initial parameters. 
To deal with this problem, we rescale the coefficients according to the methods introduced in \rcite{Shaydulin_2023, sureshbabu2023parameter}. Then the landscape after rescaling is shown in Figure~\ref{fig:landscape_rescale}, which is much easier for optimization.

\begin{figure}[htpb]
\centering
\includegraphics[scale=0.4]{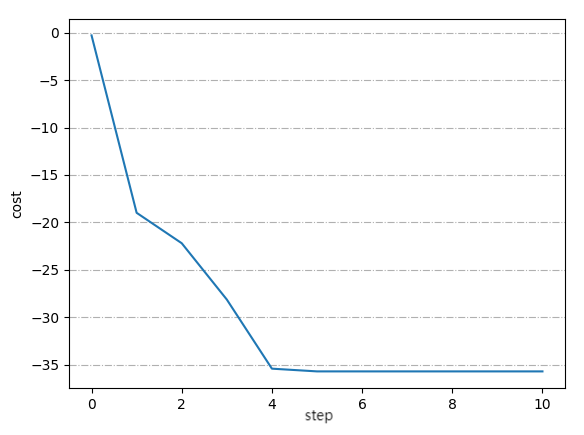}
\caption{The evolution of the expectation value of the Hamiltonian. With the iteration step increasing, the expectation value decreases and converges quickly.}
\label{fig:cost_rescale}
\end{figure}

\begin{figure}[htpb]
\centering
\includegraphics[scale=0.55]{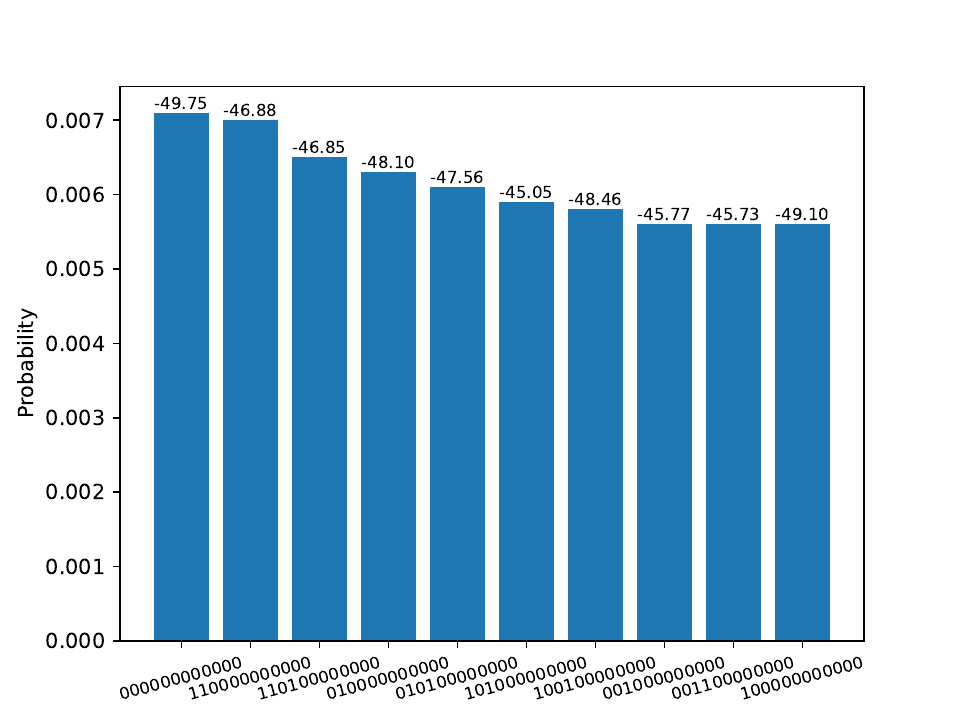}
\caption{Probability distribution for each outcome of measurements. The outcome with the highest probability corresponds to the ground state and ground energy. The ground state is all 0 with the ground energy of -49.75.}
\label{fig:prob_rescale}
\end{figure}

We use the "BFGS" optimizer for optimization and the cost function is the expectation of the HUBO Hamiltonian. The cost function during the parameter iteration process is depicted in the Figure~\ref{fig:cost_rescale}. With the iteration step increasing, the expectation value decreases and converges quickly.
After measuring the circuit for 10000 times, we obtain the probabilities of each outcome in Figure~\ref{fig:prob_rescale}. 
The outcome with the highest probability corresponds to the ground state and ground energy. Here the ground state is all 0 with the ground energy of -49.75, which is consistent to the optimal result from brute-force search.

Therefore, the simulation experiment has shown the applicability of our method on quantum computers when the number of qubits and the depth of circuit reach certain size. The quantum computation has the potential to solve such combinatorial optimization problems in the near future.

\section{Conclusions}\label{sec:conclusions}

In this study, we investigated the potential of applying quantum-inspired algorithms to solve the molecular unfolding problem. We proposed the phase encoding to encode discretized degrees of freedom, which exponentially reduces the number of binary variables for each torsional angle. We mapped the MU problem to a HUBO problem and applied the bSB algorithm for optimization. 
We used the public QM9 dataset for testing and evaluated the volume ratio over iteration step windows and the time-to-target metric. 
Compared to SA, our approach has a significant advantage in finding a solution with higher volume ratio and faster convergence. 
The median time-to-target metric is reduced by a factor of five, which demonstrates the high efficiency.
Furthermore, we showed the validity of regarding the configuration with maximal molecule volume as the most probable one by computing the RMSD between the unfolding conformation and the conformation determined by DFT. Our results showed that the median RMSD is less than about 0.5Å, indicating that solving the MU problem can provide the most probable configuration for a given molecule. This finding greatly simplifies the drug design process in real experiments.

In addition, we demonstrate a simulation experiment of MU by MindQuantum using QAOA. The optimization result from QAOA is in accordance to the optimal result, which implies our model is applicable on quantum hardware in the near future.

In the future, it is worth further studying the optimization over other degrees of freedom, including the bond lengths and bond angles. With the improvement of the objective function concerning more chemical properties, the molecular conformation generation will have more practical meanings and applications.

\section*{Acknowledgments}
Y. Li is grateful to S. Pan and J. Hu for insightful discussions.

\subsection*{Author contributions}
Yunting Li finished the main work of the paper, deduced the main formulas of the paper, plotted the figures and drafted the manuscript.
The manuscript was written and revised with contributions from all authors. 
All authors read and approved the final manuscript.

\subsection*{Funding}
B. Liu is partially supported by the National Natural Science Foundation of China (Grants No.~12101394 and No.~12171426).
The work at Fudan is supported by the National Natural Science Foundation of China (Grant No.~92165109), National Key Research and Development Program of China (Grant No.~2022YFA1404204), and Shanghai Municipal Science and Technology Major Project (Grant No.~2019SHZDZX01).

\subsection*{Data availability}
The datasets used and analysed during the current study are available from the corresponding author on reasonable request.

\section*{Declarations}

\subsection*{Ethical Approval and Consent to participate}
Not applicable.

\subsection*{Consent for publication}
Not applicable.

\subsection*{Competing interests}
The authors report no competing interests.


\end{document}


\title{Efficient molecular conformation generation with quantum-inspired algorithm: Supplemental Material}
	\maketitle

\section{Machines and libraries}

The platform for running classical algorithms is based on NUMA nodes, featuring 24 Intel(R) Xeon(R) CPU E5-2620 v3 @ 2.40GHz, Virtualization VT-x, L1d and L1i caches of 384KiB, L2 cache of 3 MiB and L3 cache of 30 MiB, RAM memory of 755G and SWAP memory of 57G. Operative system is Ubuntu 20.04.5 LTS.
The code was written in python 3.8.10. Running time of the code is counted on single CPU.

\section{One-hot encoding}\label{app:one-hot}

In the one-hot encoding method, each torsional angle $\theta_i$ corresponds to a set of $d$ binary variables $\{b_{ik}\}_k, k=0,1,\dots, d-1$. Suppose the value of each torsional angle $\theta_i$ is chosen from $d$ discrete values $\{\phi_k | k=0,1,\dots, d-1 \}$. $b_{ik}=1$ if $\theta_i=\phi_k$; otherwise $b_{ik}=0$. 
Since one torsional angle can only be assigned to one value, then the constraint of binary variables reads
\begin{align}\label{eq:one-hot constraint}
\sum_k b_{ik} = 1.
\end{align}
Similarly, this constraint holds for all variables. To embed the constraint into an unconstrained optimization problem, the following penalty term should be added to the objective function:
\begin{align}
\sum_i \left(\sum_k b_{ik} - 1 \right)^2.
\end{align}

Since the torsional angle appears in the rotation operator as $\sin(\theta_i), \cos(\theta_i)$, the expressions of these with the binary variables read \cite{Mato2022}
\begin{align}
\sin(\theta_i) = \sum_k \sin(\phi_k) b_{ik},\quad \cos(\theta_i) = \sum_k \cos(\phi_k) b_{ik}.
\end{align}
Recall that the atom positions are updated by a sequence of rotation operations defined in Eqs. (5) and (6) in the main text, then after straightforward calculations the objective function can be written as a HUBO formulation as follows:
\begin{align}
O(\{b_{ik} \}) = - D(\Theta) + A_{\mathrm{const}} \sum_i \left(\sum_k b_{ik} - 1 \right)^2,
\end{align}
where $A_{\mathrm{const}}$ is a constant for penalty strength. Then the MU problem reduces to minimizing the objective function $O(\{b_{ik} \})$.

\section{Comparison of RMSD and molecular volume in Fig.~5}

In this section, we show the values of molecular volumes and RMSDs of optimal conformations from DFT, MU, and UFF (in rdkit) in \tref{tab:mv_verify}. No. 60422 and No. 53480 are indexes of molecules with better situation in Fig.~5b, and No. 102831 and No. 50959 are indexes of molecules with worse situation in Fig.~5c in the main text. 

In some cases, the ground-truth conformation also has larger molecular volume than other conformation optimized and UFF, but in others it is not. Therefore, we compare the average behaviour of RMSD from the MU setting in the main text, and find that the molecular volume can indeed be an effective cost function for molecular conformation optimization, which greatly simplifies the conformation generation process.

\begin{table}[ht]
\caption{Molecular volumes and RMSDs of optimal conformations from DFT, MU, and UFF.}
\label{tab:mv_verify}
\centering
\renewcommand{\arraystretch}{1.5}
\begin{tabular}{@{}ccccccccc@{}}
\toprule
\multirow{2}*{\textbf{Method}} & \multicolumn{2}{c}{\textbf{No.60422}} &  \multicolumn{2}{c}{\textbf{No.53480}} & \multicolumn{2}{c}{\textbf{No.102831}} &  \multicolumn{2}{c}{\textbf{No.50959}}  \\ 
~ & RMSD & volume & RMSD & volume & RMSD & volume & RMSD & volume\\ \hline
DFT & - &  535.10 & - & 568.18 & - & 324.03 & - &  397.54\\ \hline
MU & 0.2659 & 560.81 & 0.0064 & 568.65 & 1.5341 & 395.18 & 1.0963 & 441.73 \\ \hline
UFF & 0.3801 & 546.06 & 1.0929 & 507.20 & 0.6327 & 335.82 & 0.5136 & 364.54  \\ \hline
\end{tabular}
\end{table}

\begin{figure*}[hb]
\centering
\includegraphics[scale=0.85]{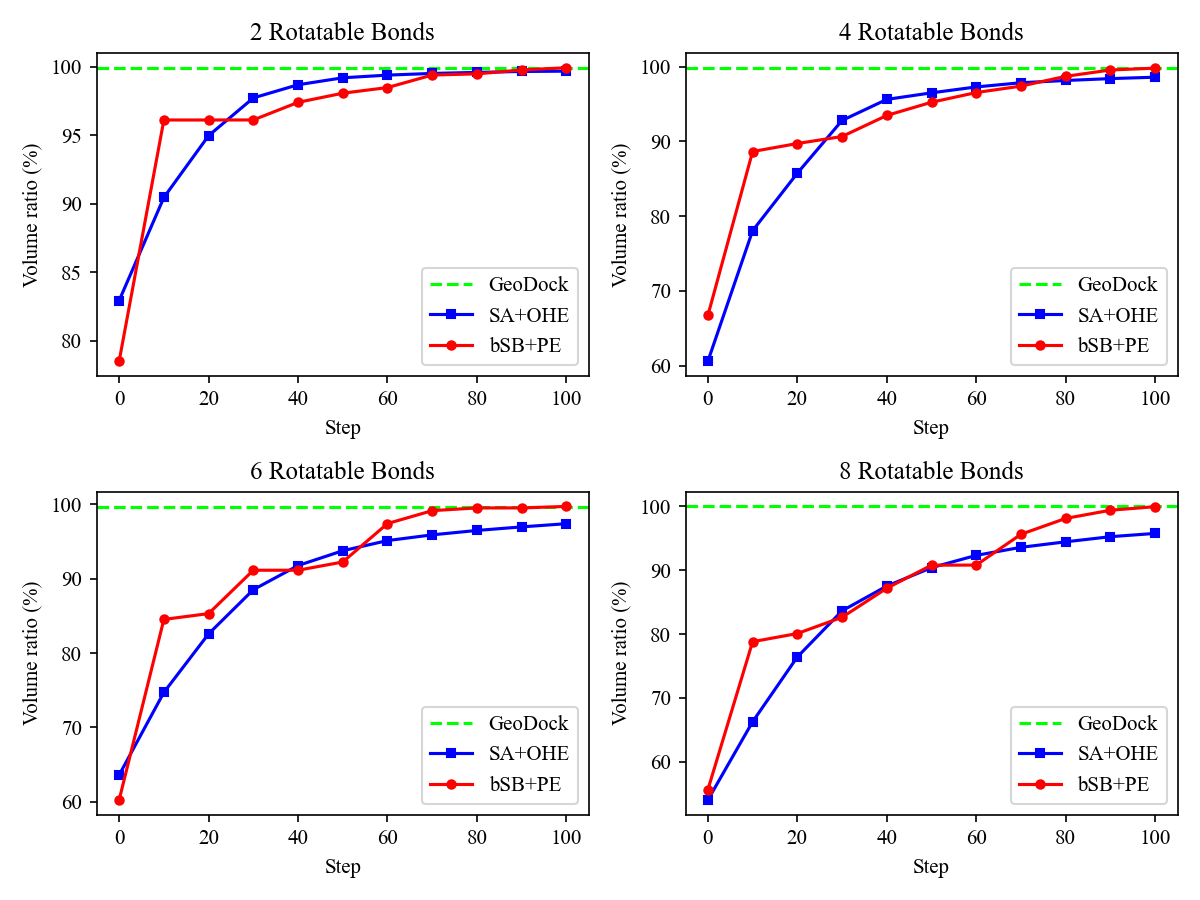}
\caption{Comparison of average volume ratio of Mato's dataset during 100 iteration steps for 2, 4, 6 and 8 rotatable bonds. Volume ratio is measured against the brute-force search of 2, 4 and 6 rotatable bonds and against greedy algorithm of 8 rotatable bonds. bSB has a faster convergence than SA at the beginning. Both SA and bSB show a good convergence to the optimal (or target) solution.}
\label{fig:accuracy}
\end{figure*}

\section{Comparison with \rcite{Mato2022}}\label{app:Dwave}

In this section, we consider the same molecules as Mato et al. considered in \rcite{Mato2022}.

\subsection{Dataset and resources estimation}

The ligand dataset is the same as Mato et al. considered in \rcite{Mato2022}. It is available in their gitlab \cite{DWave}. There are 13 molecules with a number of atoms ranging from 40 to almost 90 and a number of rotatable bonds up to 20. For comparison, we set the number of rotatable bonds up to 8.
We assume that each torsional angle is chosen from $d=16$ discrete values $\phi_k = \frac{\pi}{8} k, {k=0,1,\dots,15}$.

\subsection{Compared algorithms}

The algorithms we compared with bSB are brute-force search, GeoDock search and SA. The brute-force search gives the optimal solutions for 2, 4 and 6 rotatable bonds within bearable time costs. GeoDock search \cite{Gadioli2021} is a greedy algorithm that optimizes torsional angles one by one according to the order of betweenness centrality for each rotatable bond. It is fast but usually reaches a suboptimal solution. However, when considering 8 rotatable bonds, we use the solution from GeoDock search as a benchmark to calculate the volume ratio. For each molecule, we run the GeoDock search for 5 rounds to iteratively obtain the final solution.

\subsection{Volume ratio comparison}

In this section, we compare the volume ratio obtained by different algorithms on a fixed iteration window with an increasing number of rotatable bonds.

Volume ratio is defined against the result of brute-force search of 2, 4 and 6 rotatable bonds and against GeoDock search of 8 rotatable bonds. We show the comparison on the average volume ratio during 100 iteration steps in \fref{fig:accuracy}. In the case of 8 rotatable bonds by definition, GeoDock search reaches 100\% volume ratio and other algorithms may exceed 100\%. The best volume ratio for 8 rotatable bonds of bSB is 100.18\% while SA is 99.86\%. 
The gap at the end of iteration between SA and bSB becomes larger as the number of rotatable bonds increases. In addition, bSB has a faster convergence than SA at the beginning.

\subsection{Time-to-solution comparison}

In this section, we compare the time-to-solution (TTS) metric needed by different algorithms with an increasing number of rotatable bonds.

\begin{figure}[ht]
\centering
\includegraphics[scale=0.85]{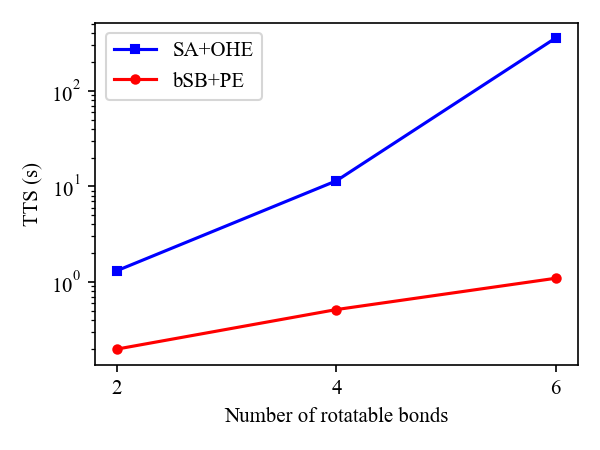}
\caption{Comparison on median TTS of Mato's dataset for 2, 4 and 6 rotatable bonds. The gap between SA and SB becomes larger as the number of rotatable bonds increases, and the TTS of SA is at least 100 times longer than that of bSB for 6 rotatable bonds.}
\label{fig:Dwave TTS}
\end{figure}

The definition of TTS is the same as Eq.(24) in the main text except that $p$ is the probability of finding the optimal solution. For each molecule, we set $N=20$ samples.
The comparsion of TTS for 2, 4 and 6 rotatable bonds after 100 iteration steps is shown in \fref{fig:Dwave TTS}. 
The gap between SA and SB becomes larger as the number of rotatable bonds increases, and the TTS of SA is at least 100 times longer than that of bSB for 6 rotatable bonds.
This result demonstrates the high efficiency of the bSB algorithm.


%